\documentclass[a4paper,11pt]{article}%
\usepackage{geometry}
\usepackage{amsmath}
\usepackage{amsfonts}
\usepackage{amssymb}
\usepackage{graphicx}
\usepackage{indentfirst}
\usepackage{bbold}
\usepackage[small,bf]{caption}
\usepackage{slashed}

\providecommand{\U}[1]{\protect\rule{.1in}{.1in}}

\newcommand{\drbar}{$\overline{\mbox{ DR}}$}
\newcommand{\ep}{\varepsilon}

\begin{document}
\date{}
\title{\textbf{Renormalization aspects of $\mathcal{N}=1$ Super Yang-Mills
    theory in the Wess-Zumino gauge}}
\author{\textbf{M.~A.~L.~Capri}$^{a}$\thanks{caprimarcio@gmail.com}\,\,,
\textbf{D.~R.~Granado}$^{a}$\thanks{diegorochagrana@uerj.br}\,\,,
\textbf{M.~S.~Guimaraes}$^{a}$\thanks{msguimaraes@uerj.br}\,\,,
\textbf{I.~F.~Justo }$^{a}$\thanks{igorfjusto@gmail.com}\,\,,\\
\textbf{L.~Mihaila }$^{b}$\thanks{luminita-nicoletta.mihaila@kit.edu}\,\,,
\textbf{S.~P.~Sorella}$^{a}$\thanks{sorella@uerj.br}\,\,\,\thanks{Work supported by
FAPERJ, Funda{\c{c}}{\~{a}}o de Amparo {\`{a}} Pesquisa do Estado do Rio de
Janeiro, under the program \textit{Cientista do Nosso Estado}, E-26/101.578/2010.}\,\,\,,
\textbf{D.~Vercauteren}\thanks{vercauteren.uerj@gmail.com}\\[2mm]
{\small \textnormal{$^{a}$  \it Departamento de F\'{\i }sica Te\'{o}rica, Instituto de F\'{\i }sica, UERJ - Universidade do Estado do Rio de Janeiro,}}
 \\ {\small \textnormal{ \it Rua S\~{a}o Francisco Xavier 524, 20550-013 Maracan\~{a}, Rio de Janeiro, Brasil}\normalsize}
 \\ {\small \textnormal{$^{b}$ \it Institut f\"{u}r Theoretische Teilchenphysik,  }}
 \\ {\small \textnormal{\phantom{$^{a}$} \it Karlsruhe Institute of Technology (KIT), D-76128 Karlsruhe, Germany}}}

\maketitle

\begin{abstract}
The renormalization of $\mathcal{N}=1$ Super Yang-Mills theory  is analysed in the Wess-Zumino gauge, employing the Landau condition. An all orders proof of the renormalizability of the theory is given by means of the Algebraic Renormalization procedure. Only three renormalization constants are needed, which can be identified with the coupling constant, gauge field and gluino renormalization. The non-renormalization theorem of the gluon-ghost-antighost vertex in the Landau gauge is shown to remain valid in $\mathcal{N}=1$ Super Yang-Mills. Moreover, due to the non-linear realization of the supersymmetry in the Wess-Zumino gauge, the renormalization factor of the gauge field turns out to be different from that of the gluino. These features are explicitly checked through a three loop calculation.

\end{abstract}

\section{Introduction}

Supersymmetric $\mathcal{N}=1$ gauge theories exhibit remarkable features, both at perturbative and non-perturbative level, see, for instance, \cite{Amati:1988ft} and refs. therein. \\\\For what concerns the ultraviolet behaviour, the symmetry between bosons and fermions gives rise to milder divergences in the ultraviolet regime, a property which is at the origin of a set of non-renormalization theorems, see  \cite{Gates:1983nr}. \\\\In this work we discuss some features of the renormalization of  $\mathcal{N}=1$ Super Yang-Mills theories in Euclidean space-time in the Wess-Zumino gauge, in which the number of field components is minimum. Employing the Algebraic Renormalization  \cite{Piguet:1995er}, we are able to show, to all orders of perturbation theory, that, in the Landau gauge, only three independent renormalization factors, $(Z_g, Z_A, Z_\lambda)$, are needed to renormalize the theory, which can be identified with the coupling constant, gauge field and gluino renormalization. The renormalization factors of all other fields, {\it i.e.} the Lagrange multiplier implementing the Landau gauge condition, the Faddeev-Popov ghosts, the external BRST sources, the global susy ghosts, etc., can be expressed as suitable combinations of $(Z_g, Z_A, Z_\lambda)$. In particular, the non-renormalization theorem of the gluon-ghost-antighost vertex in the Landau gauge, {\it i.e.} $ Z_g Z^{1/2}_A Z^{1/2}_c Z^{1/2}_{\bar c} = 1 $, still holds in $\mathcal{N}=1$ Super Yang-Mills theories, due to the existence of the so-called ghost Ward identity, see eq.\eqref{gW}. Moreover, due to the non-linear realization of suspersymmetry in the Wess-Zumino gauge, it turns out that the renormalization factor $Z_A$ of the gauge field is different from the renormalization factor $Z_\lambda$ of the gluino, a property which we shall check through a three loop calculation and which was already observed at one loop level in the Feynman gauge \cite{Hollik:2001cz}. To some extent, the present work can be seen as a continuation of the work done by \cite{White:1992ai,Maggiore:1994dw,Maggiore:1994xw,Maggiore:1995gr,Maggiore:1996gg,Ulker:2001rc, Hollik:2000pa,Hollik:2002mv,Golterman:2010zj} in which the renormalization of supersymmetric gauge theories in the Wess-Zumino gauge was faced by using BRST cohomology tools. \\\\The paper is organized as follows. In Sect.2 we discuss the BRST quantization of the theory in the Wess-Zumino gauge. In Sect.3 we derive the large set of Ward identities fulfilled by the quantized action. Further, we determine the most general invariant counterterm and find out the renormalization factors of all fields, coupling constant, and external BRST sources. Sect.4 is devoted to the explicit evaluation of the gauge field and gluino renormalization factors $(Z_A, Z_\lambda)$ as well as of the non-renormalization of the gluon-ghost-antighost vertex,  $ Z_g Z^{1/2}_A Z^{1/2}_c Z^{1/2}_{\bar c} = 1 $.  In Sect.5 we collect our conclusion. Finally, Appendix A is devoted to notations and conventions

\section{Quantization of $\mathcal N=1$ Super Yang--Mills in the Wess--Zumino gauge}
\label{sec2}
As already mentioned, the advantage of the Wess-Zumino gauge is that the number of field components is minimum. There is, however, a drawback: the supersymmetry algebra is realized in a non-linear way. More precisely, the algebra of the generators of the supersymmetry $\delta_{\alpha}$, $\alpha=1,2,3,4$, does not close on translations. Instead, we have 
\begin{equation} 
\{ \delta_\alpha, \delta_\beta \} = (\gamma_{\mu})_{\alpha\beta} \partial_{\mu} \; + \;\; ({\rm gauge \; transf.}) \; + \;\; ({\rm field \; eqs.}) \;. \label{nlalg}  
\end{equation}
As shown in \cite{White:1992ai,Maggiore:1994dw,Maggiore:1994xw,Maggiore:1995gr,Maggiore:1996gg,Ulker:2001rc}, the most powerful and efficient way to deal with the algebra \eqref{nlalg} is constructing a generalized BRST operator $Q$ which collects both gauge and susy field transformations, namely 
\begin{equation} 
Q=s+\epsilon^{\alpha}\delta_{\alpha} \;, \label{Q}
\end{equation}
where $s$ is the usual BRST operator for gauge transformations and $\epsilon^{\alpha}$ is a constant Majorana spinor parameter carrying ghost number 1. To some extent, $\epsilon^{\alpha}$ represents the ghost corresponding to the susy generators. The operator  $Q$ enjoys the following important property 
\begin{equation} 
Q^{2} = \epsilon^{\alpha}(\gamma_{\mu})_{\alpha\beta}\bar{\epsilon}^{\beta}\partial_{\mu} \;, \label{Q2}
\end{equation}
which enables us to quantize the theory by following the BRST gauge-fixing procedure in a manifestly supersymmetric invariant way. \\\\Let us proceed by showing how this construction applies to $\mathcal N=1$ Super Yang--Mills theory, whose classical action in Euclidean space\footnote{Although we are employing here the Eucliedean formulation of the theory, it is worth to point that, as far perturbation theory is concerned, the Minkowski space-time can be related to the Euclidean one through a Wick rotation. In the present paper we shall limit ourselves to perturbation theory.} reads 
\begin{equation}
\label{SYM}
S_\text{SYM} = \int d^{4}x \left[ \frac{1}{4}F^{a}_{\mu \nu}F^{a}_{\mu\nu} 
+ \frac{1}{2} \bar{\lambda}^{a\alpha} (\gamma_{\mu})_{\alpha\beta} D^{ab}_{\mu}\lambda^{b\beta}
+ \frac{1}{2}\mathfrak{D}^a\mathfrak{D}^a\right]\;,  
\end{equation}
where $D^{ab}_{\mu}= (\delta^{ab} \partial_\mu + g f^{acb}A^c_\mu)$ is the covariant derivative in the adjoint representation of the gauge group $SU(N)$,  $\lambda^{a\alpha}$ is a Majorana spinor,   $\mathfrak{D}^a$ is an auxiliary field and
\begin{equation}
F^{a}_{\mu\nu} = \partial_{\mu}A^{a}_{\nu} - \partial_{\nu}A^{a}_{\mu} + gf^{abc}A^{b}_{\mu}A^{c}_{\nu}\;.
\end{equation}

\noindent The transformation of each field under the generalised BRST operator $Q$ reads
\begin{eqnarray}
\label{susytransf}
&&
QA^{a}_{\mu} = - D^{ab}_{\mu}c^{b} 
+\bar{\epsilon}^\alpha(\gamma_\mu)_{\alpha\beta}\lambda^{a\beta}\;,\nonumber \\
&&
Q\lambda^{a\alpha} = gf^{abc}c^{b}\lambda^{c\alpha}
- \frac{1}{2}(\sigma_{\mu\nu})^{\alpha\beta}\epsilon_{\beta} F_{\mu\nu}^{a}
+ (\gamma_{5})^{\alpha\beta}\epsilon_{\beta} \mathfrak{D}^a\;, \nonumber \\
&&
Q\mathfrak{D}^a = gf^{abc}c^{b}\mathfrak{D}^c 
+ \bar{\epsilon}^{\alpha}(\gamma_{\mu})_{\alpha\beta}(\gamma_{5})^{\beta\eta}D_{\mu}^{ab}\lambda^{b}_{\eta} \;, \\
&&
Qc^{a} = \frac{1}{2}gf^{abc}c^{b}c^{c} 
- \bar{\epsilon}^{\alpha}(\gamma_{\mu})_{\alpha\beta}\epsilon^{\beta} A^{a}_{\mu}\;, \nonumber \\
&&
Q\bar{c}^{a} = b^{a}\;, \nonumber \\
&&
Qb^{a} = \nabla\bar{c}^{a} \;,\nonumber\\
&& 
Q^2= \nabla  \nonumber \;, 
\end{eqnarray}
where we have introduced the translation operator 
\begin{equation}
\label{top}
\nabla := \bar{\epsilon}^{\alpha}(\gamma_{\mu})_{\alpha\beta}\epsilon^{\beta} \partial_{\mu}\,.
\end{equation}
The fields $({\bar c}^a, c^a)$ stand for the Faddeev-Popov ghosts, while $b^a$ is the Lagrange multiplier needed to implement the Landau gauge fixing, $\partial_\mu A^a_\mu=0$. It is easy to check that the action \eqref{SYM} is left invariant by the  transformations  \eqref{susytransf}, {\it i.e.} 
\begin{equation} 
Q S_\text{SYM} = 0 \;. \label{invQ}
\end{equation}
In order to quantize the theory, we need to introduce the gauge-fixing term. This task can be accomplished by following the BRST construction, amounting to introduce the gauge condition in a $Q$-exact way. One should notice that, owing to property \eqref{Q2}, the generalised BRST operator $Q$ is in fact nilpotent when acting on space-time integrated polynomials in the fields and their derivatives. Adopting the Landau gauge, $\partial_\mu A^a_\mu=0$, for the gauge-fixing term we write  
\begin{equation}
S_\text{gf} = Q\int d^{4}x (\bar{c}^{a}\partial_{\mu}A^{a}_{\mu})\;,  \label{gfx}
\end{equation}
so that, according to \eqref{susytransf}
\begin{equation}
S_\text{gf} = \int d^{4}x \left[ \bar{c}^{a}\partial_{\mu}D^{ab}_{\mu}c^{b} 
+ b^{a}\partial_{\mu}A^{a}_{\mu} 
- \bar{c}^{a}\bar{\epsilon}^{\alpha}(\gamma_{\mu})_{\alpha\beta}\partial_{\mu}\lambda^{a\beta} \right]\;.  \label{gfx1}
\end{equation}
Therefore, the super Yang-Mills action in the Wess-Zumino and Landau gauge can be written as
\begin{eqnarray}
\label{act1}
S &=& S_{SYM} + S_\text{gf} \nonumber \\
&=&
\int d^{4}x \left\{\frac{1}{4}F^{a}_{\mu \nu}F^{a}_{\mu\nu} 
+ \frac{1}{2}\bar{\lambda}^{a\alpha}(\gamma_{\mu})_{\alpha \beta}D^{ab}_{\mu} \lambda^{b\beta}
+ \frac{1}{2}\mathfrak{D}^{2} \right. \nonumber \\
&&
\left.
+ b^{a}\partial_{\mu}A^{a}_{\mu}
+\bar{c}^{a}\left[\partial_{\mu}D^{ab}_{\mu}c^{b} - \bar{\epsilon}^{\alpha}(\gamma_{\mu})_{\alpha \beta}\partial_{\mu}\lambda^{a\beta} \right]\right\}.
\label{gfaction}
\end{eqnarray}
From eqs.\eqref{susytransf}, \eqref{invQ}, \eqref{gfx}, it follows immediately that 
\begin{equation}
Q S = 0 \;, \label{invS}
\end{equation}
meaning that the gauge fixing procedure has been done in a BRST invariant way. Moreover, reminding that  the generalized operator $Q$ collects both gauge and supersymmetry transformations, one realizes that the expression \eqref{gfx1} is the supersymmetric generalization of the Landau gauge, as it can be inferred from the presence of the additional term $\bar{c}^{a}\bar{\epsilon}^{\alpha}(\gamma_{\mu})_{\alpha\beta}\partial_{\mu}\lambda^{a\beta}$, which contains the supersymmetry ghost $\bar{\epsilon}^{\alpha}$ as well as the gluino field $\lambda^{a\beta}$. \\\\Having quantized the theory, we are ready to write down the large set of Ward identities and proceed with the algebraic characterization of the most general invariant counterterm. This will be the task of the next section.

\section{Ward identities and algebraic characterization of the invariant counterterm}
In order to write down the set of Ward identities which will be employed for the algebraic analysis of the model, we need to introduce a set of external sources coupled to the non-linear transformations appearing in eqs.\eqref{susytransf}. More precisely, from \eqref{susytransf}, we need to introduce external sources coupled to $QA^a_\mu$, $Q\lambda^{a\beta}$, $QD^a$ and $Qc^a$.  To that purpose, we introduce the following BRST doublets \cite{Piguet:1995er} of sources
\begin{equation}
\left\{\begin{matrix}QK^{a}_{\mu}=\Omega^{a}_{\mu}\phantom{\Bigl|}\cr
Q\Omega^{a}_{\mu}=\nabla K^{a}_{\mu}\phantom{\Bigl|}\end{matrix}\right.\,,\qquad
\left\{\begin{matrix}QL^{a}=\Lambda^{a}\phantom{\Bigl|}\cr
Q\Lambda^{a}=\nabla L^{a}\phantom{\Bigl|}\end{matrix}\right.\,,\qquad
\left\{\begin{matrix}QT^{a}= J^{a}\phantom{\Bigl|}\cr
QJ^{a}=\nabla T^{a}\phantom{\Bigl|}\end{matrix}\right.\,,\qquad
\left\{\begin{matrix}QY^{a\alpha}=X^{a\alpha}\phantom{\Bigl|}\cr
QX^{a\alpha}=\nabla Y^{a\alpha}\phantom{\Bigl|}\end{matrix}\right.\,,
\end{equation}
and the $Q-$exact external action 
\begin{equation}
S_\text{ext} = Q \int d^{4}x \left( -K^{a}_{\mu} A^{a}_{\mu} + L^{a}c^{a} - T^{a} \mathfrak{D}^a  + Y^{a\alpha}\lambda^{a}_{\alpha}  - T^a Y^{a\alpha} (\gamma_5)_{\alpha \beta} \epsilon^{\beta} \right)\;,  \label{qexact}
\end{equation}
leading to the following complete $Q-$invariant action $\Sigma$ 
\begin{equation}
\Sigma = S_{SYM} + S_\text{gf} + S_\text{ext}   \;, \label{complact}
\end{equation}
\begin{equation}
Q \Sigma = 0 \;. 
\end{equation}
Explicitly 
\begin{eqnarray}
\label{fullact}
\Sigma &=& \int d^{4}x \biggl\{ \frac{1}{4}F^{a}_{\mu \nu}F^{a}_{\mu\nu}
+ \frac{1}{2} \bar{\lambda}^{a\alpha}(\gamma_{\mu})_{\alpha \beta}D^{ab}_{\mu}\lambda^{b\beta}
+ \frac{1}{2}\mathfrak{D}^{a}\mathfrak{D}^{a}
+ b^{a}\partial_{\mu}A^{a}_{\mu}
\nonumber \\
&&
+\bar{c}^{a}\Bigl[\partial_{\mu}D^{ab}_{\mu}c^{b}
-\bar{\epsilon}^{\alpha}(\gamma_{\mu})_{\alpha \beta}\partial_{\mu}\lambda^{a\beta}\Bigr]
+ T^{a}\Bigl[gf^{abc}c^{b}\mathfrak{D}^{c}
+ \bar{\epsilon}^{\alpha}(\gamma_{\mu})_{\alpha \beta}(\gamma_{5})^{\beta \eta} D_{\mu}^{ab}\lambda^{b}_{\eta}\Bigr]
\nonumber \\
&&
+L^{a}\Bigl[ \frac{g}{2}f^{abc}c^{b}c^{c}
-\bar{\epsilon}^\alpha(\gamma_{\mu})_{\alpha\beta}\epsilon^\beta A^{a}_{\mu}\Bigr]
- K^{a}_{\mu}\Bigl[D^{ab}_{\mu}c^{b}
- \bar{\epsilon}^\alpha(\gamma_\mu)_{\alpha\beta}\lambda^{a\beta}\Bigr]
- \Omega^{a}_{\mu} A^{a}_{\mu}
\nonumber\\
&&
+Y^{a\alpha}\Bigl[ gf^{abc}c^{b}\lambda^{c}_{\alpha} - \frac{1}{2}(\sigma_{\mu\nu})_{\alpha\beta} F_{\mu\nu}^{a}\epsilon^{\beta}
+ (\gamma_{5})_{\alpha\beta}\epsilon^{\beta} \mathfrak{D}^a\Bigr]
+ \Lambda^{a}c^{a}
- J^{a}\mathfrak{D}^{a}
+ X^{a\alpha}\lambda^{a}_{\alpha}
\nonumber\\
&&
-J^{a}Y^{a\alpha}(\gamma_{5})_{\alpha\beta}\varepsilon^{\beta}
+T^{a}X^{a\alpha}(\gamma_{5})_{\alpha\beta}\varepsilon^{\beta}
\biggl\}\,.
\end{eqnarray}
Notice that in expression \eqref{qexact} a term quadratic in the external sources, {\it i.e.}$T^a Y^{a\alpha} (\gamma_5)_{\alpha \beta} \epsilon^{\beta}$,  has been introduced. Similar terms are present also in the analysis done by \cite{White:1992ai,Maggiore:1994dw,Maggiore:1994xw,Maggiore:1995gr,Maggiore:1996gg,Ulker:2001rc}. As we shall see, it will be needed for  renormalization purposes. The external sources can be set to zero at the end, after having identified the most general counter term and all renormalization factors. Expression \eqref{fullact} represents the starting point for the algebraic analysis of the model, namely for the determination of the most general invariant counterterm compatible with all possible Ward identities fulfilled by $\Sigma$.

\subsection{Ward identities }

The complete action $\Sigma$ obeys a large set of Ward identities, which we display below:  

\begin{itemize}
{\item The Slavnov-Taylor identity:}

\begin{equation}
\mathcal{S}(\Sigma) = 0 \;, \label{STid}
\end{equation}
where 
\begin{eqnarray}
\mathcal{S}(\Sigma) &=& \int d^{4}x \biggl\{\biggl(\frac{\delta \Sigma}{\delta A^{a}_{\mu}}
+ \Omega^{a}_{\mu}\biggr)\frac{\delta \Sigma}{\delta K^{a}_{\mu}}
+ \biggl(\frac{\delta \Sigma}{\delta \lambda^{a\alpha}}
+ X^{a\alpha}\biggr)\frac{\delta \Sigma}{\delta Y^{a\alpha}}
+ \frac{\delta \Sigma}{\delta \lambda^{a\alpha}}(\gamma_{5})_{\alpha\beta}\epsilon^{\beta}J^{a}
\nonumber \\
&&
+\biggl(\frac{\delta \Sigma}{\delta c^{a}}
+ \Lambda^{a}\biggr)\frac{\delta \Sigma}{\delta L^{a}} 
+ \biggl(\frac{\delta \Sigma}{\delta \mathfrak{D}^{a}}
+ {J}^{a}\biggr)\frac{\delta \Sigma}{\delta T^{a}}
- \frac{\delta \Sigma}{\delta \mathfrak{D}^{a}}X^{a\alpha}(\gamma_{\alpha\beta})\epsilon^{\beta}
+ b^{a}\frac{\delta \Sigma}{\delta \bar{c}^{a}}
\nonumber \\
&&
+ (\nabla \bar{c}^{a})\frac{\delta \Sigma}{\delta b^{a}}
+(\nabla K^{a}_{\mu})\frac{\delta \Sigma}{\delta \Omega^{a}_{\mu}}
+(\nabla Y^{a\alpha})\frac{\delta \Sigma}{\delta {X}^{a\alpha}}
+(\nabla T^{a})\frac{\delta \Sigma}{\delta {J}^{a}}
+(\nabla L^{a})\frac{\delta \Sigma}{\delta {\Lambda}^{a}}\biggr\}
\nonumber\,.
\label{ST}
\end{eqnarray}
From the Slavnov-Taylor identity  \eqref{STid}, it follows that the so-called linearized operator $\mathcal{B}_{\Sigma}$ \cite{Piguet:1995er}

\begin{eqnarray}
\mathcal{B}_{\Sigma} &=& \int d^{4}x \biggl\{ 
\frac{\delta \Sigma}{\delta K^{a}_{\mu}}\frac{\delta }{\delta A^{a}_{\mu}}
+ \frac{\delta \Sigma}{\delta A^{a}_{\mu}}\frac{\delta }{\delta K^{a}_{\mu}}
+ \Omega^{a}_{\mu}\frac{\delta }{\delta K^{a}_{\mu}}
+ \frac{\delta \Sigma}{\delta Y^{a\alpha}}\frac{\delta }{\delta \lambda^{a\alpha}}
\nonumber \\
&&
+ \frac{\delta \Sigma}{\delta \lambda^{a\alpha}}\frac{\delta }{\delta Y^{a\alpha}}
+ X^{a\alpha} \frac{\delta }{\delta Y^{a\alpha}}
+ \frac{\delta}{\delta \lambda^{a\alpha}}(\gamma_{5})_{\alpha\beta}\epsilon^{\beta}J^{a}
+ \frac{\delta \Sigma}{\delta L^{a}} \frac{\delta }{\delta c^{a}}
+ \frac{\delta \Sigma}{\delta c^{a}} \frac{\delta }{\delta L^{a}}
\nonumber \\
&&
+ \Lambda^{a} \frac{\delta }{\delta L^{a}} 
+ \frac{\delta \Sigma}{\delta T^{a}} \frac{\delta }{\delta \mathfrak{D}^{a}}
+ \frac{\delta \Sigma}{\delta \mathfrak{D}^{a}} \frac{\delta }{\delta T^{a}}
+ {J}^{a} \frac{\delta }{\delta T^{a}}
- X^{a\alpha}(\gamma_{\alpha\beta})\epsilon^{\beta}\frac{\delta}{\delta \mathfrak{D}^{a}}
+ b^{a}\frac{\delta }{\delta \bar{c}^{a}}
\nonumber \\
&&
+ (\nabla \bar{c}^{a})\frac{\delta }{\delta b^{a}}
+(\nabla K^{a}_{\mu})\frac{\delta }{\delta \Omega^{a}_{\mu}}
+(\nabla Y^{a\alpha})\frac{\delta }{\delta {X}^{a\alpha}}
+(\nabla T^{a})\frac{\delta }{\delta {J}^{a}}
+(\nabla L^{a})\frac{\delta }{\delta {\Lambda}^{a}}\biggr\}  \;, 
\label{LST}
\end{eqnarray}
enjoys the following property 
\begin{equation}
\mathcal{B}_{\Sigma}  \mathcal{B}_{\Sigma}  = \nabla \;, \label{bnilp}
\end{equation}
so that $\mathcal{B}_{\Sigma}$ is nilpotent when acting on integrated functionals.  
\item{The Landau gauge-fixing condition and the anti-ghost equation \cite{Piguet:1995er}:}
\begin{equation}
\frac{\delta\Sigma}{\delta b^{a}}= \partial_{\mu}A^{a}_{\mu}\,,\qquad
\frac{\delta\Sigma}{\delta\bar{c}^{a}}+\partial_{\mu}\frac{\delta\Sigma}{\delta K^{a}_{\mu}}=0\,.
\label{GFandAntiGhost}
\end{equation}

\item{The ghost Ward identity \cite{Blasi:1990xz,Piguet:1995er}:}
\begin{equation}
G^{a}(\Sigma)=\Delta^{a}_{\mathrm{class}}\,,  \label{gW}
\end{equation}
where
\begin{equation}
G^{a}:=\int d^{4}x\,\biggl[\frac{\delta}{\delta{c}^{a}} 
+ gf^{abc}\bar{c}^{b}\frac{\delta}{\delta{b}^{c}}\biggl]\,,
\end{equation}
and
\begin{equation}
\Delta^{a}_{\mathrm{class}}=\int d^{4}x\,\left[gf^{abc}\left(K^{b}_{\mu}A^{c}_{\mu}
-L^{b}c^{c}+T^{b}\mathfrak{D}^{a}
-Y^{b\alpha}\lambda^{c}_{\alpha}\right)-\Lambda^{a}\right]\,.
\end{equation}
Notice that the breaking term $\Delta^{a}_{\mathrm{class}}$ appearing in the right-hand side of eq.\eqref{gW} is linear in the quantum fields. As such, $\Delta^{a}_{\mathrm{class}}$ is a classical breaking, not affected by quantum corrections \cite{Blasi:1990xz,Piguet:1995er}.

\item{The equation of motion of the auxiliary field $\mathfrak{D}^{a}$:}

\begin{equation}
\frac{\delta\Sigma}{\delta{\mathfrak{D}^{a}}}=  \mathfrak{D}^{a} 
- J^{a} + gf^{abc}c^{b}T^{c} + Y^{a\alpha}(\gamma_{5})_{\alpha\beta}\,\varepsilon^{\beta}\,.  \label{auxW}
\end{equation}
Again, being linear in the quantum fields, the right-hand side of  \eqref{auxW} is a classical breaking. 

\item{The linearly broken gluino Ward identity, namely:}

\begin{equation}
\label{eqT}
\left[ \frac{\delta}{\delta{T^{a}}}
+ (\gamma_{5})_{\alpha\beta}\,\varepsilon^{\beta}\frac{\delta}{\delta{\lambda^{a}_{\alpha}}}
+gf^{abc}\left( c^{b}\frac{\delta }{\delta{\mathfrak{D}^{c}}}
- T^{b}\frac{\delta}{\delta{L^{c}}} 
\right)\right]\Sigma = \tilde\Delta^{a}_{\mathrm{class}}
\end{equation}
where  $\tilde\Delta^{a}_{\mathrm{class}}$ is a classical breaking 
\begin{eqnarray}
\tilde{\Delta}^{a}_{\mathrm{class}} &=&
3gf^{abc}\bar{\epsilon}^{\alpha}(\gamma_{\mu})_{\alpha\beta}\epsilon^{\beta} T^{b}A^{c}_{\mu}
+ \nabla T^{b}
-gf^{abc}c^{b}J^{c} \nonumber \\
&&
+ \bar{\epsilon}^{\alpha}(\gamma_{\mu})_{\alpha\eta}(\gamma_{5})^{\eta\beta}\epsilon_{\beta} \left( \partial_{\mu}\bar{c}^{a} + K^{a}_{\mu} \right) \;.
\end{eqnarray}

\end{itemize}
We notice, in particular,  that the gluino Ward identity \eqref{eqT} follows by commuting the Slavnov-Taylor identity \eqref{STid} with equation \eqref{auxW}. \\\\Before turning to the algebraic analysis of the most general invariant counterterm, let us spend a few words on the role of the auxiliary fields  $\mathfrak{D}^{a}$, which we have introduced in the expression of the starting action  $S_\text{SYM}$, eq.\eqref{SYM}. As it is apparent from eq.\eqref{SYM}, the fields $\mathfrak{D}^{a}$ enter the action  $S_\text{SYM}$ only quadratically. As such, they do not play any role in the loop calculations. Though,  they are needed in order to write down the Slavnov-Taylor identities \eqref{STid}, which are at the basis of the Algebraic Renormalization   set up \cite{Piguet:1995er}. Here, we have two equivalent options. The first option is that of starting from the beginning by including the $\mathfrak{D}^{a}$ fields  in the action, eq.\eqref{SYM}, as well as in the $Q$-transformations \eqref{susytransf}. In this case, the BRST operator $Q$ enjoys the important property 
\begin{equation}
Q^2 = \nabla  \;, \label{qdn}
\end{equation}
which enables us to construct the Slavnov-Taylor identities in the way described before. The second option is that of not including the fields $\mathfrak{D}^{a}$ from the beginning, see, for instance, \cite{Maggiore:1994dw,Maggiore:1994xw,Maggiore:1995gr}. This means that the fields $\mathfrak{D}^{a}$ are absent in both the starting action as well as the $Q$-transformations. However, the BRST operator $Q$ does not display now the property \eqref{qdn}. Instead, one has
\begin{equation}
Q^2 = \nabla + {\rm eqs.\; of\; motion}  \;. \label{qdn1}
\end{equation}
In this case, in order to establish the Slavnov-Taylor identities, an additional care has to be taken. The presence of terms proportional to the equations of motion in eq.\eqref{qdn1} requires the introduction of terms which are quadratic in the BRST sources \cite{Maggiore:1994dw,Maggiore:1994xw,Maggiore:1995gr}. These terms are precisely of the same kind of $\mathfrak{D}^{a} \mathfrak{D}^{a}$. At the end of this second procedure, one is able to write down Slavnov-Taylor identities which are exactly of the same type of  \eqref{STid} \cite{Maggiore:1994dw,Maggiore:1994xw,Maggiore:1995gr}, so that both options give the same results for the characterization of the invariant counterterm.

\subsection{Algebraic characterization of the invariant counterterm and renormalizability of the $\mathcal{N}=1$ Super-Yang-Mills}
In order to determine the most general invariant counterterm which can be freely added to each order, we follow the Algebraic Renormalization framework  \cite{Piguet:1995er} and perturb  the complete action $\Sigma$ by adding an integrated local polynomial in the fields and sources with dimension four and vanishing ghost number, $\Sigma_{count}$, and we require that the perturbed action, $(\Sigma + \omega \Sigma_{count})$, where $\omega$ is an infinitesimal expansion parameter, obeys the same Ward identities fulfilled by $\Sigma$ to the first order in the parameter $\omega$, namely 
\begin{equation} 
\mathcal{S}(\Sigma + \omega \Sigma_{count}) = 0 + O(\omega^2)\;, \label{stp}
\end{equation}
\begin{equation}
\frac{\delta(\Sigma+ \omega \Sigma_{count})}{\delta b^{a}}= \partial_{\mu}A^{a}_{\mu} + O(\omega^2) \,,\qquad
\left( \frac{\delta}{\delta\bar{c}^{a}}+\partial_{\mu}\frac{\delta} {\delta K^{a}_{\mu}}\right) (\Sigma+ \omega \Sigma_{count}) =0+ O(\omega^2) \,,
\label{GFandAntiGhostp}
\end{equation}
\begin{equation}
G^{a}(\Sigma+ \omega \Sigma_{count})=\Delta^{a}_{\mathrm{class}}+O(\omega^2)\,,  \label{gWp}
\end{equation}
\begin{equation}
\frac{\delta(\Sigma+ \omega \Sigma_{count})}{\delta{\mathfrak{D}^{a}}}=  \mathfrak{D}^{a} 
- J^{a} + gf^{abc}c^{b}T^{c} + Y^{a\alpha}(\gamma_{5})_{\alpha\beta}\,\varepsilon^{\beta}+ O(\omega^2)\,,  \label{auxWp}
\end{equation}
\begin{equation}
\label{eqTp}
\left[ \frac{\delta}{\delta{T^{a}}}
+ (\gamma_{5})_{\alpha\beta}\,\varepsilon^{\beta}\frac{\delta}{\delta{\lambda^{a}_{\alpha}}}
+gf^{abc}\left( c^{b}\frac{\delta }{\delta{\mathfrak{D}^{c}}}
- T^{b}\frac{\delta}{\delta{L^{c}}} 
\right)\right](\Sigma + \omega \Sigma_{count}) = \tilde\Delta^{a}_{\mathrm{class}} + O(\omega^2) \;. 
\end{equation}
To the first order in the expansion parameter $\omega$, equations  \eqref{stp}, \eqref{GFandAntiGhostp},  \eqref{gWp}, \eqref{auxWp}, \eqref{eqTp} give rise to the following constraints: 
\begin{equation}
\mathcal{B}_{\Sigma} (\Sigma_{count}) = 0\;,  \label{c1}
\end{equation}
\begin{equation}
\frac{\delta}{\delta b^{a}}\,\Sigma_{count}=0 \;, \qquad  \left( \frac{\delta}{\delta\bar{c}^{a}} + \partial_\mu\frac{\delta}{\delta K_\mu^{a}} \right) \Sigma_{count}=0\;, \label{c2}
\end{equation}
\begin{equation}
G^{a}\,\Sigma_{count}=0  \;, \label{c3}
\end{equation}
\begin{equation}
\frac{\delta}{\delta \mathfrak{D}^{a}}\,\Sigma_{count}=0  \;, \label{c4}
\end{equation}
and 
\begin{equation}
\label{c5}
\left[ \frac{\delta}{\delta{T^{a}}}
+ (\gamma_{5})_{\alpha\beta}\,\varepsilon^{\beta}\frac{\delta}{\delta{\lambda^{a}_{\alpha}}}
+gf^{abc}\left( c^{b}\frac{\delta }{\delta{\mathfrak{D}^{c}}}
- T^{b}\frac{\delta}{\delta{L^{c}}} 
\right)\right]\Sigma_{count} =  0 \;, 
\end{equation}
where $\mathcal{B}_{\Sigma}$ stands for the linearized operator of eq.\eqref{LST}. 
The first condition, eq.\eqref{c1}, tells us that $\Sigma_{count}$ belongs to the cohomology of the operator $\mathcal{B}_{\Sigma}$ in the space of the local integrated polynomials in the fields and external sources of dimension bounded by four. From the general results on the cohomology of Yang-Mills theories, see \cite{Piguet:1995er} and refs. therein, it follows that $\Sigma_{count}$ can be parametrized as follows
\begin{equation}
\label{cttnt}
\Sigma_{count}=a_{0}\,S_{\mathrm{SYM}} + \mathcal{B}_{\Sigma} \Delta^{(-1)} \;.
\end{equation}
where $a_0$ is a free coefficient and $\Delta^{(-1)}$ stands for the most general integrated local polynomial in the fields and sources, with ghost number $-1$ and dimension $3$. 
\begin{center}
\begin{table}
\label{Table1}
\begin{tabular}{l|c|c|c|c|c|c|c|c|c|c|c|c|c|c|c|c}
&$A$&$\lambda$&$\mathfrak{D}$&$c$&$\bar{c}$&$b$&$K$&$\Omega$&$\Lambda$&$T$&$J$&$L$&$Y$&$X$&$
\epsilon$&$\bar{\epsilon}$\cr
\hline\hline
$\phantom{\Bigl|}\!\!$Dim
&1&$\frac{3}{2}$&2&1&1&2&2&3&3&1&2&2&$\frac{3}{2}$&$\frac{5}{2}$&$\frac{1}{2}$&$\frac{1}{2}$\cr
\hline
$\phantom{\Bigl|}\!\!$Ghost\#
&0&0&0&1&$-1$&0&$-1$&0&$-1$&$-1$&0&$-2$&$-1$&0&1&1\cr
\hline
$\phantom{\Bigl|}\!\!$Nature
&C&A&C&A&A&C&A&C&A&A&C&C&C&A&C&C
\end{tabular}
\caption{Quantum numbers of all fields and sources. "A" stands for anti-commuting, while "C" for commuting.}
\end{table}
\end{center}
From Table 1, the most general expression for $\Delta^{(-1)}$ can be written as
\begin{eqnarray}
\Delta^{(-1)}&=&\int d^{4}x\,\biggl\{
a_{1}\,  \partial_{\mu}\bar{c}^{a}A^{a}_{\mu}
+a_{2}\, K^{a}_{\mu}A^{a}_{\mu}
+a_{3}\, T^{a}\,\partial_{\mu}A^{a}_{\mu}
+a_{4}\, b^{a}\bar{c}^{a}
+a_{5}\, b^{a}T^{a}
+a_{6}\, \mathfrak{D}^{a}\,\bar{c}^{a}
\nonumber\\
&&
+a_{7}\, J^{a}T^{a}
+a_{8}\, \lambda^{a\alpha}Y_{a\alpha}
+a_{9}\, Y^{a\alpha}\,(\gamma_{5})_{\alpha\beta}\epsilon^{\beta}T^{a}
+a_{10}\, gf^{abc}\,\bar{c}^{a}\bar{c}^{b}c^{c}
+a_{11}\, J^{a}\bar{c}^{a}
\nonumber\\
&&
+a_{12}\, \bar{c}^{a}\,\epsilon^{\alpha}\,(\gamma_{5})_{\alpha\beta}\,Y^{a\beta}
+a_{13}\, gf^{abc}T^{a}T^{b}c^{c}
+a_{14}\, \mathfrak{D}^{a}\,T^{a}
\nonumber\\
&&
+a_{15}\, gf^{abc}\,c^{a}\bar{c}^{b}T^{c}
+a_{16}\, c^{a}L^{a}\biggr\}\,,
\end{eqnarray}
with $a_{i}$ ($i = 1$ to $16$) being arbitrary coefficients. It is worth to point out that, according to Table 1, the ultraviolet dimension of both ghost and anti-ghost fields, $(c, {\bar c})$, has been chosen to be equal to 1. This feature turns out to be very helpful, as enables us to assign positive ultraviolet dimension $1/2$ to the supersymmetric parameter $\epsilon$, a property which greatly simplifies the analysis of the invariant counterterm  $\Sigma_{count}$. \\\\From eqs.\eqref{c2}, \eqref{c3}, \eqref{c4}, \eqref{c5}, it follows that \begin{equation}
a_{1}=a_{2}\;, \qquad a_{14} = -\frac{a_{0}}{2}\;, \qquad a_{9}= \left( \frac{a_{0}}{2} - a_{8}\right) \qquad \text{and}
\end{equation}
\begin{eqnarray}
&&
a_{3}=a_{4}=a_{5}=a_{6}=a_{7}=a_{10}=a_{11}=a_{12} =0 \;, \nonumber \\ 
&&
a_{13}=a_{15}=a_{16}=a_{17}=a_{18}=a_{19}=0\;,
\end{eqnarray}
leading to 
\begin{equation}
\Delta^{(-1)} = \int d^{4}x\, \biggl\{ 
a_{1} (\partial_{\mu}\bar{c}^{a} + K^{a}_{\mu})A^{a}_{\mu}
+ a_{8}Y^{a\alpha}\lambda^{a}_{\alpha}
+ \left(\frac{a_{0}}{2}-a_{8}\right)Y^{a\alpha}(\gamma_{5})_{\alpha\beta}\epsilon^{\beta}T^{a}
- \frac{a_{0}}{2} \mathfrak{D}^{a} T^{a}
\biggl\}\;.
\end{equation}
Therefore, for the exact part of expression \eqref{cttnt}, {\it i.e.}  $\mathcal{B}_{\Sigma} \Delta^{(-1)}$, we get 
\begin{eqnarray}
\mathcal{B}_{\Sigma}\Delta^{(-1)} &=& \int d^{4}x \biggl\{
a_{1}\left(\frac{\delta \Sigma}{\delta A^{a}_{\mu}} + \Omega^{a}_{\mu} +\partial_{\mu}b^{a}\right)A^{a}_{\mu}
+ a_{8}\left(\frac{\delta \Sigma}{\delta\lambda^{a}_{\alpha}}+X^{a\alpha}\right)\left(\lambda^{a}_{\alpha}-
(\gamma_{5})_{\alpha\beta}\epsilon_{\beta}T^{a}\right) \nonumber\\
&&
+ \frac{a_{0}}{2}\left(\frac{\delta\Sigma}{\delta\lambda^{a}_{\alpha}}+X^{a\alpha}\right)(\gamma_{5})_{\alpha\beta}\epsilon^{\beta}T^{a}
- \frac{a_{0}}{2}\left(\frac{\delta\Sigma}{\delta \mathfrak{D}^{a} }+J^{a}\right)\left(\mathfrak{D}^{a} -Y^{a\alpha}(\gamma_{5})_{\alpha\beta}\epsilon^{\beta}\right) \nonumber \\
&&
- a_{8}\left(\frac{\delta\Sigma}{\delta \mathfrak{D}^{a} }+J^{a}\right)Y^{a\alpha}(\gamma_{5})_{\alpha\beta}\epsilon^{\beta}
- a_{1}\bar{c}^{a}\frac{\delta\Sigma}{\delta\bar{c}^{a}}
- a_{1}K^{a}_{\mu}\frac{\delta\Sigma}{\delta K^{a}_{\mu}}
+ a_{8}Y^{a\alpha}\frac{\delta\Sigma}{\delta Y^{a\alpha}} \nonumber \\
&&
+ \frac{a_{0}}{2}T^{a}\frac{\delta\Sigma}{\delta T^{a}} \bigg\}\;.
\label{ct}
\end{eqnarray}
yielding the final form of the most general invariant counterterm
\begin{eqnarray}
\Sigma_{count} &=& \int d^{4}x \left\{
\frac{a_{0}}{4}F^{a}_{\mu\nu}F^{a}_{\mu\nu}
+ a_{1}\frac{\delta \Sigma_{SYM}}{\delta A^{a}_{\mu}}A^{a}_{\mu}
+ \frac{(a_{0} - 2a_{8})}{2}\bar{\lambda}^{a\alpha}(\gamma_{\mu})_{\alpha\beta}D^{ab}_{\mu}\lambda^{b\beta} \right. \nonumber \\
&&
+ a_{1}\left( \partial_{\mu}\bar{c}^{a} + K^{a}_{\mu}\right)\partial_{\mu}c^{a}
+ (a_{1} + a_{8}) \bar{\epsilon}^{\alpha}(\gamma_{\mu})_{\alpha\beta}\lambda^{a\beta}\left(\partial_{\mu}\bar{c}^{a} + K^{a}_{\mu}\right) \nonumber \\
&&
+ (a_{0} - 2a_{8})\bar{\epsilon}^{\alpha}(\gamma_{\mu})_{\alpha\beta}(\gamma_{5})^{\beta\eta}T^{a}D^{ab}_{\mu}\lambda^{b}_{\eta}
- a_{1}gf^{abc}T^{a}\bar{\epsilon}^{\alpha}(\gamma_{\mu})_{\alpha\beta}(\gamma_{5})^{\beta\eta}\lambda^{b}_{\eta}A^{c}_{\mu}\nonumber \\
&&
- a_{1}\bar{\epsilon}^{\alpha}(\gamma_{\mu})_{\alpha\beta}\epsilon^{\beta}A^{a}_{\mu}L^{a}
+ \left( a_{8} - \frac{a_{0}}{2}\right)\bar{\epsilon}^{\alpha}(\gamma_{\mu})_{\alpha\beta}\epsilon^{\beta}T^{a}D^{ab}_{\mu}T^{b} \nonumber \\
&&
+ \left( \frac{a_{0}}{2}-a_{8}\right) \left(Y^{a\alpha}(\gamma_{5})_{\alpha\beta}\epsilon^{\beta}\right)^{2}
- \frac{1}{2}(a_{1}+a_{8})Y^{a\alpha}(\sigma_{\mu\nu})_{\alpha\beta}\epsilon^{\beta}\left(\partial_{\mu}A^{a}_{\nu} - \partial_{\nu}A^{a}_{\mu}\right) \nonumber \\
&&
\left.
-\left(a_{1}+\frac{a_{8}}{2}\right) gf^{abc} Y^{a\alpha}(\sigma_{\mu\nu})_{\alpha\beta}\epsilon^{\beta}A^{b}_{\mu}A^{c}_{\nu} \right\}\;.
\label{lastct}
\end{eqnarray}
One sees that $\Sigma_{count}$ contains three arbitrary coefficients, $a_0, a_1, a_8$, which will identify the renormalization factors of all fields, sources and coupling constant. To complete the analysis of the algebraic renormalization of the model, we need to show that the counterterm $\Sigma_{count}$ can be reabsorbed into the starting action $\Sigma$ through a redefinition of the fields and parameters $\{\phi \}$ , $\phi = (A,\lambda, b, c, \bar c, \mathfrak{D}, \epsilon)$,  of the sources $\{ S \}$, $S= (K,\Omega, \Lambda, T, J, L, Y, X )$,  and coupling constant $g$, namely 
\begin{equation}
\label{ration}
\Sigma(\phi,S,g) + \omega \Sigma_{count}(\phi,S,g)  = \Sigma(\phi_0,S_0,g_0) + O(\omega^2) \;, 
\end{equation}
where $(\phi_0, S_0, g_0)$ stand for the so-called bare fields, sources and coupling constant:
\label{renormfs}
\begin{equation}
\phi_{0}=Z^{1/2}_{\phi}\,\phi  \qquad\;,   \qquad
S_{0}=Z_{S}\,S\,,  \qquad g_0 = Z_g g   \;, 
\end{equation}
and the renormalization factors  $Z$ can be written as
\begin{equation}
Z^{1/2}_{\phi}=(1+\omega\,z_\phi)^{1/2}=1+\omega \frac{z_{\phi}}{2}+O(\omega^{2})\,,\qquad
Z_{S}=1+\omega\,z_S\,, \qquad Z_g = 1 +\omega z_g \;.  
\end{equation}
Moreover, in the present case, a little care has to be taken with the potential mixing of quantities which have the same quantum numbers. In fact, from equation \eqref{ct} one can easily notice  that the field $\lambda^{a\alpha}$ and the combination $\gamma_{5}\epsilon T^{a}$ have the same dimension and quantum numbers as well as the field $\mathfrak{D}^{a}$ and the combination $Y^{a}\gamma_{5}\epsilon$, as it can be checked from Table 1.   As a consequence, these quantities can mix at the quantum level, a well known property of renormalization theory. This feature can be properly taken into account by writing the renormalization of the fields $\lambda$ and $\mathfrak{D}$ in matrix form, {\it i.e.}  
\begin{equation}
\label{lrenorm}
\lambda^{a\alpha}_{0}=Z^{1/2}_{\lambda}\,\lambda^{a\alpha}+\omega\, z_{1}\,T^{a}(\gamma_{5})^{\alpha\beta}\varepsilon_{\beta}
\end{equation}
and
\begin{equation}
\label{drenorm}
\mathfrak{D}^{a}_{0}=Z^{1/2}_{\mathfrak{D}}\,\mathfrak{D}^{a}+\omega\, z_{2}\,Y^{a\alpha}(\gamma_{5})_{\alpha\beta}\varepsilon^{\beta}\,,
\end{equation}
while the remaining fields, sources and parameters still obey \eqref{renormfs}.\\\\From direct inspection of equation \eqref{ration}, the renormalization factors of all fields, sources and parameters are given by 
\begin{eqnarray}
Z_A^{1/2}&=&1+\omega\left(\frac{a_{0}}{2}+a_1\right)\,, \nonumber \\
Z_{g}&=&1-\omega\frac{a_{0}}{2}\,, \nonumber \\
Z_{\lambda}^{1/2}&=&1+\omega\left(\frac{a_{0}}{2}-a_{8}\right)\,,  \label{three} 
\end{eqnarray}
while the remaining factors are
\begin{eqnarray}
Z_T&=&Z_{g}^{-1/2}Z_{A}^{1/4}\,, \nonumber \\
Z_{\varepsilon}&=&Z_{g}^{1/2}Z_{A}^{-1/4}\,, \nonumber \\
Z_{Y}&=&Z_{g}^{-1/2}Z_{A}^{1/4}Z_{\lambda}^{-1/2}\,,\nonumber \\
Z_{b}^{1/2}&=&Z_{A}^{-1/2}\,,\nonumber \\
Z_{L}&=&Z_{A}^{1/2}\,,\nonumber \\
Z_{c}^{1/2}&=&Z_{\bar{c}}^{1/2}=Z_{K}=Z_{g}^{-1/2}Z_{A}^{-1/4}\,,\nonumber \\
Z_{\Lambda}&=&Z_{g}^{1/2}Z_{A}^{1/4}\,,\nonumber \\
Z_{J}&=&1\,,\nonumber \\
Z_{\mathfrak{D}}&=&1\,,\nonumber \\
Z_{X}&=&Z_{\lambda}^{-1/2}\,, \nonumber \\
Z_{\Omega}&=&Z_{A}^{-1/2}  \label{other}
\label{eq:wid}
\end{eqnarray}
and
\begin{equation}
z_1=-z_2=a_8-\frac{a_0}{2}\,.  \label{z1z2}
\end{equation}
We have thus completed the all order proof of the algebraic renormalization of $\mathcal{N}=1$ supersymmetric Yang-Mills theories. A few remarks are in order. Three independent parameters, $a_0, a_1, a_8$, are needed to renormalize the theory. According to eqs.\eqref{three}, these parameters correspond to the renormalization of the gauge coupling constant $g$, of the gauge field $A^a_\mu$ and of the gluino $\lambda^{a\alpha}$. The renormalization constants of all other fields, sources and parameters can be written down as suitable  combinations of $Z_g, Z_A, Z_\lambda$, as expressed by eqs.\eqref{other},\eqref{z1z2}. We remark that the celebrated nonrenormalization theorem of the gluon- ghost-antighost vertex of the Landau gauge \cite{Blasi:1990xz,Piguet:1995er}, {\it i.e.} $Z_{c}^{1/2}Z_{\bar{c}}^{1/2}Z_{g}Z_{A}^{1/2}=1$, remains valid in the supersymmetric version of the theory. Moreover, although belonging to the same multiplet, eqs.\eqref{three} suggest that the renormalization constant of the gauge field, $Z_A$, turns out to be different from that of the gluino, $Z_\lambda$. That this will be in fact the case, will be shown in the next section, where the explicit three loop expression of $Z_A,Z_\lambda$ will be reported.

\section{Three-loop calculation of the renormalization factors $Z_A$ and $Z_\lambda$ and check of the non-renormalization theorem of the gluon-ghost-antighost vertex}
We explicitly computed the wave-function renormalization constants for
the  bosonic and fermionic degrees of freedom  $Z_A$, $Z_\lambda$, and $Z_c$
and the gauge coupling renormalization constant $Z_g$ up to
three loops in perturbation theory. As renormalization scheme we used the
minimal  subtraction scheme with dimensional reduction~\cite{siegel} (DRED) as
regulator. Such  renormalization scheme is commonly denoted as  \drbar{}.
Let us mention that we  applied DRED in the component field formalism and
implemented its mathematical consistent formulation~\cite{Avdeev:1981vf,ds}. It is well known
that DRED in this formulation  breaks supersymmetry
in higher orders of perturbation theory~\cite{Avdeev:1982xy}. Nevertheless, for a Supersymmetric
Yang-Mills theory it has been proven explicitely that DRED preserves 
supersymmetry up to three loops~\cite{Harlander:2006xq,Jack:2007ni}. \\\\The advantage of this scheme is that all ultraviolet (UV) counterterms
are polynomial both in external momenta and
masses~\cite{Collins:1974bg}. The most effective approach
is its use in combination with
multiplicative renormalization. This amounts in general to
solve recursively the equation
\begin{eqnarray}
Z_a &=& 1-K_\ep[\Gamma_a(p^2) Z_a]\,,
\label{eq::rec}
\end{eqnarray}
where $K_\ep [f(\ep)]$ stands for the singular part of the Laurent
expansion of $f(\ep)$ in $\ep$ around $\ep=0$.  $\Gamma_a(p^2)$
denotes the renormalized Green function with only one external momentum $p^2$
kept non-zero.  $Z_a$ denotes the renormalization constant associated
with the Green function  $\Gamma_a$. In this case, the renormalization
of $\Gamma_a$  through $(l+1)$-loop
order requires  the
renormalization of the Lagrangian parameters like couplings, masses,
gauge parameters, etc. up to $l$-loop order. For the present calculation we
considered the renormalization of the Green functions corresponding to the
gauge boson propagator, its ghost and its Majorana superpartner propagators
and the vertices  containing ghost-gauge boson and Majorana fermion-gauge
boson interactions. \\\\For the explicit calculation of Feynman diagrams up to three-loop order, we used a well-tested chain of programs:
 {\tt QGRAF}~\cite{Nogueira:1991ex} generates all contributing Feynman
 diagrams. The 
output is passed via {\tt
  q2e}~\cite{Harlander:1997zb,Seidensticker:1999bb}, which transforms
Feynman diagrams into 
Feynman amplitudes, to {\tt exp}~\cite{Harlander:1997zb,Seidensticker:1999bb} 
that generates {\tt FORM}\cite{Vermaseren:2000nd} code. The latter is
processed by {\tt   MINCER}~\cite{Larin:1991fz} 
which computes analytically  massless propagator diagrams up to three
loops  and outputs the  $\epsilon$ expansion of the result. Here,
$ \epsilon= (4-d)/2$  is the regulator of Dimensional Regularization with $d$ being the space-time dimension used for
the evaluation of the momentum integrals. \\\\We performed all the calculations in a linear gauge and only in the last step
specified the results to the Landau gauge. This procedure  allows us to check
explicitly the gauge independence of the gauge coupling renormalization
constant. In our setup, the gauge
parameter $\xi$ is defined through the gauge boson propagator
\begin{eqnarray}
D^A_{\mu\nu}&=& -i \frac{g_{\mu\nu} - (1-\xi)\frac{q_\mu q_\nu}{q^2}}{q^2+i\varepsilon}\,.
\end{eqnarray}
The three-loop expression for the wave-function renormalization constant of
the Majorana field   reads
\begin{eqnarray}
Z_{\lambda} &=& 1-\frac{1}{\epsilon}\left(\frac{\alpha}{4\pi}\right)C_{A}\xi 
+\left( \frac{\alpha}{4\pi}\right)^{2}C_{A}^{2}\left[\frac{1}{4\epsilon^{2}}3\xi(1+\xi)
-\frac{1}{8\epsilon}(3+8\xi+\xi^{2}) \right] \nonumber \\
&&
+\left(\frac{\alpha}{4\pi}\right)^{3}C_{A}^{3}\left[- \frac{1}{8\epsilon^{3}}\xi(9+9\xi+4\xi^{2})
+\frac{1}{4\epsilon^{2}}(3+11\xi+7\xi^{2}+\xi^{3}) \right. \nonumber \\
&&
\left. +\frac{1}{96\epsilon}(66-108 Z_3 -3\xi(53+8 Z_3) -3\xi^2(13+4 Z_3)
-10\xi^3 )
\right]\;.
\label{eq::gluino}
\end{eqnarray}
Here $Z_{3}=\zeta_{3}$ is the Riemann $\zeta$-function,
$\alpha=\frac{g}{4\pi}$ and $C_{A}$ is the quadratic Casimir invariant in the
adjoint representation. In the special case of the  Landau gauge, for which
$\xi=0$, it reduces to 
\begin{eqnarray}
Z_{\lambda} &=& 1-
\left( \frac{\alpha}{4\pi}\right)^{2} \frac{3}{8\epsilon}C_{A}^{2}
+\left(\frac{\alpha}{4\pi}\right)^{3}C_{A}^{3}\left[+\frac{3}{4\epsilon^{2}}
  +\frac{1}{48\epsilon}(33-54 Z_{3})\right]\;.  \label{gluino_landau}
\end{eqnarray}
For the three-loop expression of the wave-function renormalization constant of
the gauge boson, we obtained
\begin{eqnarray}
Z_{A} &=& 1+\left(\frac{\alpha}{4\pi}\right)\frac{1}{2\epsilon}C_{A}(3-\xi) 
+\left(\frac{\alpha}{4\pi}\right)^{2}C_{A}^{2}\left[\frac{1}{8\epsilon^{2}}(-9-3\xi+2\xi^{2})
\frac{1}{16\epsilon}(27-11\xi-2\xi^{2})  \right]\nonumber \\
&&
+\left(\frac{\alpha}{4\pi}\right)^{3}C_{A}^{3}\left[\frac{1}{16\epsilon^{3}}(27+9\xi-2\xi^{3})
+\frac{1}{96\epsilon^{2}}(-369-39\xi+60\xi^{2}+14\xi^{3})  \right. \nonumber \\
&&
\left. +\frac{1}{96\epsilon}\left(533-7\xi^{3}-114 Z_{3}-3\xi^{2}(11+2 Z_{3}) 
-\xi(113+24 Z_{3})\right)\right]\;.
\label{eq::gluon}
\end{eqnarray}
It is an easy exercise to obtain its expression for the Landau gauge 
\begin{eqnarray}
Z_{A} &=& 1+ \left(\frac{\alpha}{4\pi}\right)\frac{3 C_{A}}{2 \epsilon}
+\left(\frac{\alpha}{4\pi}\right)^{2}C_{A}^{2}\left[-\frac{9}{8\epsilon^{2}}+\frac{27}{16\epsilon}\right] 
\nonumber \\
&&
+\left(\frac{\alpha}{4\pi}\right)^{3}C_{A}^{3}\left[\frac{27}{16\epsilon^{3}}-\frac{123}{32\epsilon^{2}}+\frac{1}{96\epsilon}(533-114Z_{3})\right]\;.
\label{eq::gluon2}
\end{eqnarray}
The expression for the three-loop wave function renormalization constant of the
ghost is given by
\begin{eqnarray}
Z_{c} &=& 1+\left(\frac{\alpha}{4\pi}\right)\frac{1}{4\epsilon}C_{A}(3-\xi) 
+\left(\frac{\alpha}{4\pi}\right)^{2}C_{A}^{2}\left[\frac{3}{32\epsilon^{2}}(-9+\xi^{2})+
\frac{1}{32\epsilon}(21+\xi)  \right]\nonumber \\
&&
+\left(\frac{\alpha}{4\pi}\right)^{3}C_{A}^{3}\left[\frac{1}{128\epsilon^{3}}(189+9\xi-9\xi^2-5\xi^{3})
+\frac{1}{384\epsilon^{2}}(-891+12\xi+39\xi^{2}+8\xi^{3})  \right. \nonumber \\
&&
\left. +\frac{1}{192\epsilon}\left(139-3\xi^{3}+114 Z_{3}+6\xi^{2}(-1+ Z_{3}) 
+24\xi Z_{3})\right)\right]\;.
\label{eq::ghost}
\end{eqnarray}
The simplified formula for the case of the Landau gauge reads
 \begin{eqnarray}
Z_{c} &=& 1+\left(\frac{\alpha}{4\pi}\right)\frac{3}{4\epsilon}C_{A} 
+\left(\frac{\alpha}{4\pi}\right)^{2}C_{A}^{2}\left[-\frac{27}{32\epsilon^{2}}+
\frac{21}{32\epsilon}  \right]
\nonumber \\
&&
+\left(\frac{\alpha}{4\pi}\right)^{3}C_{A}^{3}\left[\frac{189}{128\epsilon^{3}}
-\frac{297}{128\epsilon^{2}}
+\frac{1}{192\epsilon}\left(139+114 Z_{3} \right)\right]\;.
\label{eq::ghost2}
\end{eqnarray}
Our results for the three-loop renormalization constant of the gauge coupling
completely agree with the previous calculations of Refs.~\cite{Avdeev:1981ew,Jack:1996vg,Jack:2007ni}.
For convenience of the reader we quote them  below
 \begin{eqnarray}
Z_{g} &=& 1-\left(\frac{\alpha}{4\pi}\right)\frac{3}{2\epsilon}C_{A} 
+\left(\frac{\alpha}{4\pi}\right)^{2}C_{A}^{2}\left[\frac{27}{8\epsilon^{2}}-
\frac{3}{2\epsilon}  \right]
+\left(\frac{\alpha}{4\pi}\right)^{3}C_{A}^{3}\left[-\frac{135}{16\epsilon^{3}}
+\frac{33}{4\epsilon^{2}}
-\frac{7}{2\epsilon}\right]\;.
\label{eq::gauge}
\end{eqnarray}
Using eqs.~(\ref{eq::gluon2}),(\ref{eq::ghost2}),(\ref{eq::gauge})  one can immediately test the non-renormalization of the gluon-ghost-antighost vertex, given in eqs.(\ref{eq:wid}), {\it i.e.} $Z_g Z^{1/2}_A Z_c=1$.

\section{Conclusion}
In this work the issue of the renormalization of $\mathcal{N}=1$ Super Yang-Mills theory has been addressed in the Wess-Zumino gauge, by employing the Landau condition. Following the setup already outlined by the authors  \cite{White:1992ai,Maggiore:1994dw,Maggiore:1994xw,Maggiore:1995gr,Maggiore:1996gg,Ulker:2001rc}, the renormalization of the theory has been investigated within the Algebraic Renormalization framework \cite{Piguet:1995er}, through BRST cohomology tools. \\\\Our main result is summarized by eqs.\eqref{three},\eqref{eq:wid}. In the Landau gauge, only three renormalization factors, $Z_g, Z_A, Z_\lambda$, are needed in order to renormalize the theory. The renormalization constants of all other fields can be expressed as suitable combinations of $Z_g, Z_A, Z_\lambda$, as displayed by eqs.\eqref{eq:wid}. Moreover, although belonging to the same multiplet, the renormalization constant of the gauge field, $Z_A$, turns out to be different from that of the gluino, $Z_\lambda$, as explicitly checked through the three loop computations, see eqs.\eqref{gluino_landau},\eqref{eq::gluon2}. As already mentioned, this  feature is due to the use of the Wess-Zumino gauge, in which the supersymmetry is realized in a non-linear way. Further, the non-renormalization theorem of the gluon-ghost-antighost vertex has been shown to remain valid in $\mathcal{N}=1$ Super Yang-Mills.\\\\Finally, although we have limited ourselves to consider only the case of pure $\mathcal{N}=1$ Super Yang-Mills theory, the inclusion of matter fields can be done straightforwardly. Let us also point out that the non-renormalization of the gluon-ghost-antighost vertex remains valid in presence of matter fields, as a consequence  of the ghost Ward identity, eq.\eqref{gW}, which still holds in presence of matter \cite{Piguet:1995er,Maggiore:1994dw,Maggiore:1994xw,Maggiore:1995gr}.

\section*{Acknowledgments}
The Conselho Nacional de Desenvolvimento Cient\'{\i}fico e
Tecnol\'{o}gico (CNPq-Brazil), the Faperj, Funda{\c{c}}{\~{a}}o de
Amparo {\`{a}} Pesquisa do Estado do Rio de Janeiro, the Latin
American Center for Physics (CLAF), the SR2-UERJ,  the
Coordena{\c{c}}{\~{a}}o de Aperfei{\c{c}}oamento de Pessoal de
N{\'{\i}}vel Superior (CAPES)  are gratefully acknowledged. 
LM is supported by the DFG through SFB/TR 9.
\begin{appendix}
\section{Notations and conventions in Euclidean space-time}
\label{notations}
\noindent
\textbf{Units}: $\hbar=c=1$.

\noindent
\textbf{Euclidean metric}: $\delta_{\mu\nu}=diag(+,+,+,+)$.

\noindent
\textbf{Wick rotations:} $X_0\rightarrow -iX_4\Rightarrow\partial_0\rightarrow+i\partial_4$, $A_0\rightarrow+iA_4$

\noindent
\textbf{Gamma matrices:}
\[\gamma_4=\left( \begin{array}{cc}
0 & \mathbb{1} \\
\mathbb{1} & 0 \end{array} \right),~~
\gamma_k=-i\left( \begin{array}{cc} 0 & \sigma_k \\ -\sigma_k & 0 \end{array} \right)\]

\noindent
\textbf{Pauli matrices}:
\[\sigma_4=\left( \begin{array}{cc}
1 & 0 \\
0 & 1 \end{array} \right),~~
\sigma_1=\left( \begin{array}{cc} 0 & 1\\ 1 & 0 \end{array} \right),~~
\sigma_2=\left( \begin{array}{cc} 0 & -i\\ i & 0 \end{array} \right),~~
\sigma_3=\left( \begin{array}{cc} 1 & 0 \\ 0 & -1 \end{array} \right)\]

\noindent
The Gamma matrices obey the following properties:
\begin{eqnarray}
 \gamma_\mu=\gamma_\mu^\dagger\\
  \{\gamma_\mu,\gamma_\nu\}&=&2\delta_{\mu\nu}
\end{eqnarray}

\noindent
We also define the $\gamma_5$ matrix as:
\[\gamma_5=\gamma_4\gamma_1\gamma_2\gamma_3=\left(\begin{array}{cc}
                                                   \mathbb{1} & 0\\
                                                   0 & -\mathbb{1}\\
                                                  \end{array}\right)\]

\noindent
with the following properties:
\begin{equation}
 \{\gamma_5,\gamma_\mu\}=0,~~(\gamma_5)^2=\mathbb{1},~~\gamma_5^\dagger=\gamma_5
\end{equation}

\noindent
The charge conjugation matrix is:
\begin{equation}
\label{Cmtrx}
\mathcal{C}=\gamma_4\gamma_2=i\left(\begin{array}{cc} \sigma_2 & 0 \\ 0 & -\sigma_2\\ \end{array}\right)
\end{equation}

\noindent
with the following properties:
\begin{equation}
 \mathcal{C}^{-1}=-\mathcal{C}=\mathcal{C}^T,~~\mathcal{C}^{-1}\gamma_\mu\mathcal{C}=-
\gamma_\mu^T
\end{equation}

\noindent
The $\sigma^{\mu\nu}$ tensor is defined as
\begin{equation}
  (\sigma_{\mu\nu})^{~\beta}_\alpha\equiv\frac{1}{2}[\gamma_\mu,\gamma_\nu]^{~\beta}_\alpha
\end{equation}
and has the property $\sigma_{\mu\nu}^\dagger=-\sigma_{\mu\nu}$.

\noindent
\textbf{Majorana fermions:}

\noindent
The Majorana condition reads:
\begin{equation}
\label{mjconj}
 \lambda^\mathcal{C}=\lambda=\mathcal{C}\bar{\lambda}^T~~
\Longleftrightarrow~~\bar{\lambda}=\lambda^T\mathcal{C}\;,
\end{equation}
leading to the following relations
\begin{equation}
\bar{\lambda}\gamma_{\mu}\epsilon = \bar{\epsilon}\gamma_{\mu}\lambda \qquad \text{and} \qquad \bar{\lambda}\gamma_{\mu}\gamma_{5}\epsilon = - \bar{\epsilon}\gamma_{\mu}\gamma_{5}\lambda\;.
\end{equation}

\noindent
\textbf{Fierz identity (in Euclidean space-time):}
\begin{eqnarray}
\epsilon_{1}\bar{\epsilon}_{2} &=&  \frac{1}{4}(\bar{\epsilon}_{2}\epsilon_{1})\mathbb{1} 
+ \frac{1}{4}(\bar{\epsilon}_{2}\gamma_{5}\epsilon_{1})\gamma_{5}
+ \frac{1}{4}(\bar{\epsilon}_{2}\gamma_{\mu}\epsilon_{1})\gamma_{\mu}
- \frac{1}{4}(\bar{\epsilon}_{2}\gamma_{\mu}\gamma_{5}\epsilon_{1})\gamma_{\mu}\gamma_{5}\nonumber \\
&&
- \frac{1}{8}(\bar{\epsilon}_{2}\sigma_{\mu\nu}\epsilon_{1})\sigma_{\mu\nu}    \;.
\end{eqnarray}

\noindent {\bf Indices notations}:
\begin{center}
\begin{tabular}{ll}
$\bullet$&The Lorentz indices: $\mu,\nu,\rho,\sigma,\lambda\in\{1,2,3,4\}$\,;$\phantom{\Bigl|}$\\
$\bullet$&The Spinor indices: $\alpha,\beta,\gamma,\delta,\eta\in\{1,2,3,4\}$\,;$\phantom{\Bigl|}$\\
$\bullet$&The $SU(N)$ group indices: $a,b,c,d,e\in\{1,\dots,N^{2}-1\}$\,;$\phantom{\Bigl|}$\\
\end{tabular}
\end{center}

\end{appendix}


\begin{thebibliography}{9}


\bibitem{Amati:1988ft} 
  D.~Amati, K.~Konishi, Y.~Meurice, G.~C.~Rossi and G.~Veneziano,
  Phys.\ Rept.\  {\bf 162}, 169 (1988).
  
\bibitem{Gates:1983nr} 
  S.~J.~Gates, M.~T.~Grisaru, M.~Rocek and W.~Siegel,
  Front.\ Phys.\  {\bf 58}, 1 (1983)
  [hep-th/0108200].


\bibitem{Piguet:1995er}
  O.~Piguet and S.~P.~Sorella,
  Lect.\ Notes Phys.\ M {\bf 28} (1995) 1.
  
\bibitem{Hollik:2001cz} 
  W.~Hollik and D.~Stockinger,
  Eur.\ Phys.\ J.\ C {\bf 20}, 105 (2001)
  [hep-ph/0103009].
  
   
\bibitem{White:1992ai}
  P.~L.~White,
  Class.\ Quant.\ Grav.\  {\bf 9} (1992) 1663.

  
\bibitem{Maggiore:1994dw} 
  N.~Maggiore,
  Int.\ J.\ Mod.\ Phys.\ A {\bf 10}, 3781 (1995)
  [hep-th/9501057].
  
\bibitem{Maggiore:1994xw} 
  N.~Maggiore,
  Int.\ J.\ Mod.\ Phys.\ A {\bf 10}, 3937 (1995)
  [hep-th/9412092].
  
  
  
\bibitem{Maggiore:1995gr} 
  N.~Maggiore, O.~Piguet and S.~Wolf,
  Nucl.\ Phys.\ B {\bf 458}, 403 (1996)
  [Erratum-ibid.\ B {\bf 469}, 513 (1996)]
  [hep-th/9507045].
  
  
\bibitem{Maggiore:1996gg}
  N.~Maggiore, O.~Piguet and S.~Wolf,
  Nucl.\ Phys.\ B {\bf 476} (1996) 329
  [hep-th/9604002].
  
\bibitem{Ulker:2001rc}
  K.~Ulker,
  Mod.\ Phys.\ Lett.\ A {\bf 17} (2002) 739
  [hep-th/0108062].
  
\bibitem{Blasi:1990xz} 
  A.~Blasi, O.~Piguet and S.~P.~Sorella,
  Nucl.\ Phys.\ B {\bf 356}, 154 (1991).
  
   
\bibitem{Hollik:2000pa} 
  W.~Hollik, E.~Kraus and D.~Stockinger,
  Eur.\ Phys.\ J.\ C {\bf 23}, 735 (2002)
  [hep-ph/0007134].
  
\bibitem{Hollik:2002mv} 
  W.~Hollik, E.~Kraus, M.~Roth, C.~Rupp, K.~Sibold and D.~Stockinger,
  Nucl.\ Phys.\ B {\bf 639}, 3 (2002)
  [hep-ph/0204350].
  
\bibitem{Golterman:2010zj} 
  M.~Golterman and Y.~Shamir,
  Phys.\ Rev.\ D {\bf 82}, 105003 (2010)
  [arXiv:1004.3860 [hep-th]].
  
  

\bibitem{siegel}
 W.~Siegel,
 Phys.\ Lett.\ B {\bf 84} (1979) 193.

\bibitem{Avdeev:1981vf}
  L.V.~Avdeev, G.A.~Chochia and A.A.~Vladimirov,
  Phys.\ Lett.\ B {\bf 105} (1981) 272.



\bibitem{ds}  
  D.~St\"ockinger,
  JHEP {\bf 0503} (2005) 076
  [arXiv:hep-ph/0503129].


\bibitem{Avdeev:1982xy}
  L.~V.~Avdeev and A.~A.~Vladimirov,
  Nucl.\ Phys.\  B {\bf 219} (1983) 262.

\bibitem{Harlander:2006xq}
  R.~V.~Harlander, D.~R.~T.~Jones, P.~Kant, L.~Mihaila and M.~Steinhauser,
  JHEP {\bf 0612} (2006) 024
  [arXiv:hep-ph/0610206].


\bibitem{Jack:2007ni}
  I.~Jack, D.~R.~T.~Jones, P.~Kant and L.~Mihaila,
  JHEP {\bf 0709} (2007) 058
  [arXiv:0707.3055 [hep-th]].


\bibitem{Collins:1974bg}
  J.~C.~Collins,
  Nucl.\ Phys.\  B {\bf 80} (1974) 341; Nucl.\ Phys.\  B {\bf 92} (1975) 477.



\bibitem{Nogueira:1991ex}
  P.~Nogueira,
  J.\ Comput.\ Phys.\  {\bf 105} (1993) 279.



\bibitem{Harlander:1997zb}
  R.~Harlander, T.~Seidensticker and M.~Steinhauser,
  Phys.\ Lett.\ B {\bf 426} (1998) 125
  [hep-ph/9712228].



\bibitem{Seidensticker:1999bb}
  T.~Seidensticker,
  [hep-ph/9905298].


\bibitem{Vermaseren:2000nd}
  J.~A.~M.~Vermaseren,
  arXiv:math-ph/0010025.


\bibitem{Larin:1991fz}
  S.~A.~Larin, F.~V.~Tkachov and J.~A.~M.~Vermaseren,
preprint NIKHEF-H-91-18 (1991).

\bibitem{Avdeev:1981ew}
  L.~V.~Avdeev and O.~V.~Tarasov,
  Phys.\ Lett.\  B {\bf 112}, 356 (1982).


\bibitem{Jack:1996vg}
  I.~Jack, D.~R.~T.~Jones and C.~G.~North,
  Phys.\ Lett.\  B {\bf 386} (1996) 138
  [arXiv:hep-ph/9606323].


  \end{thebibliography}
\end{document}